\documentclass[runningheads]{llncs}
\usepackage{graphicx}
\usepackage{multirow}
\usepackage{amssymb}
\usepackage{caption}
\usepackage{subcaption}
\begin{document}
\title{TransFusion: Multi-view Divergent Fusion for Medical Image Segmentation with Transformers}

%
\author{
Di Liu\inst{1} 
\and
Yunhe Gao\inst{1} 
\and
Qilong Zhangli\inst{1} 
\and
Ligong Han\inst{1} 
\and
Xiaoxiao He\inst{1} 
\and
Zhaoyang Xia\inst{1} 
\and
Song Wen\inst{1} 
\and
Qi Chang\inst{1} 
\and
Zhennan Yan\inst{2} 
\and
Mu Zhou\inst{2} 
\and
Dimitris Metaxas\inst{1} 
}
\authorrunning{D. Liu et al.}
%
\institute{
Department of Computer Science, Rutgers University \\
\and
SenseBrain Research
\\
}
\maketitle              
\begin{abstract}
Combining information from multi-view images is crucial to improve the performance and robustness of automated methods for disease diagnosis. However, due to the non-alignment characteristics of multi-view images, building correlation and data fusion across views largely remain an open problem. In this study, we present TransFusion, a Transformer-based architecture to merge divergent multi-view imaging information using convolutional layers and powerful attention mechanisms. In particular, the Divergent Fusion Attention (DiFA) module is proposed for rich cross-view context modeling and semantic dependency mining, addressing the critical issue of capturing long-range correlations between unaligned data from different image views. We further propose the Multi-Scale Attention (MSA) to collect global correspondence of multi-scale feature representations. We evaluate TransFusion on the Multi-Disease, Multi-View \& Multi-Center Right Ventricular Segmentation in Cardiac MRI (M\&Ms-2) challenge cohort. TransFusion demonstrates leading performance against the state-of-the-art methods and opens up new perspectives for multi-view imaging integration towards robust medical image segmentation.

\end{abstract}
\section{Introduction}
\begin{figure*} [t]
\begin{center}
\includegraphics[width=1\textwidth]{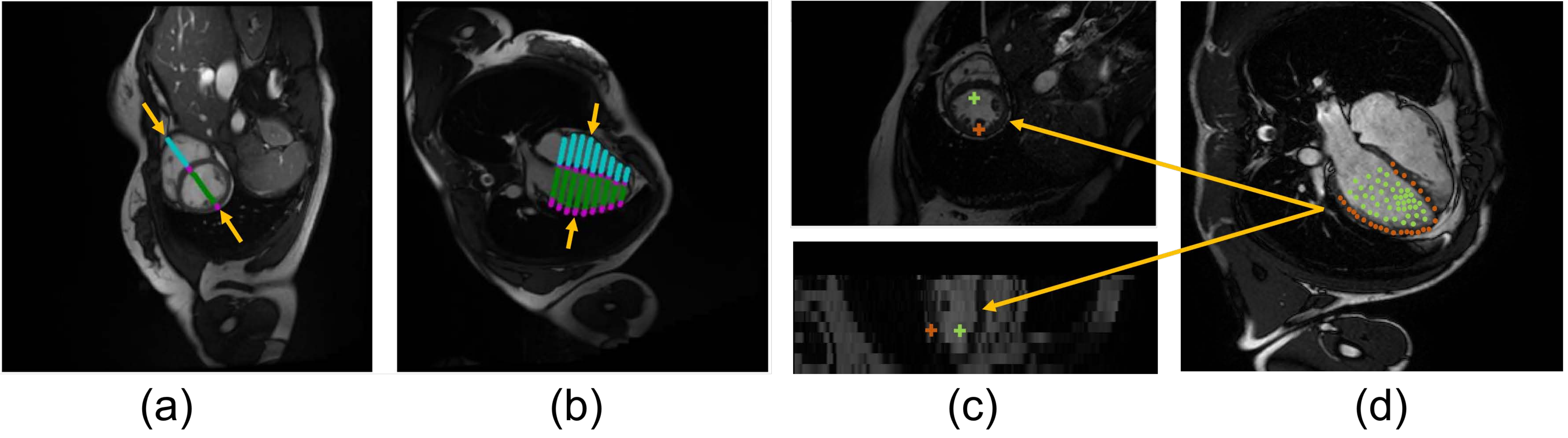}
\end{center}
   \caption{(a) and (b): The correspondence of ventricular annotation between the short-axis and long-axis views. (c) and (d) show an example of how the cross markers in (c) aggregate features from corresponding dots in (d) via learned attention map using DiFA module. The bottom image in (c) is the sagittal view of short-axis image stack with low resolution. }
\label{example}
\end{figure*}

Multi-view medical image analysis allows us to combine the strengths from different views towards fully understanding of cardiac abnormalities, wall motion, and outcome diagnosis in clinical workflows~\cite{xia2020uncertainty,wang2017multi,tian2018cr,vigneault2018omega,hu2020harnessing,zhao2019applications,liu2021label}. For instance, cardiac MRI (cMRI) given appropriate MR acquisition settings can  produce images with high intra-slice resolution and low inter-slice resolution \cite{petitjean2011review,liu2019dispersion,remedios2021joint,he2019effective}. Short-axis cMRI analysis is prone to fail in the base and the apex \cite{bernard2018deep,hu2003vivo,chang2020soft} due to low inter-slice resolution and possible misalignment. Many of these issues and missing information can be resolved with long-axis cMRI scans~\cite{campello2021multi,ge2020automated,liu2020dispersion,chang2022deeprecon}. Nevertheless, finding corresponding structures and fusing multi-view information is challenging, because the locations of the regions of interest in different views can be relatively far apart (see Fig. \ref{example}). Due to this distinct structure unalignment among views, long-range correspondence modeling is highly desired. Convolution on simply concatenated unaligned multi-view images lacks the ability to find  such correspondences
\cite{wang2017multi,li2021right}.




Transformers with the self-attention mechanism are gaining momentum in  medical image analysis~\cite{dosovitskiy2020image,gao2021utnet,hatamizadeh2022unetr,cao2021swin,gao2022multi,zhangli2022region}. The transformer self-attention mechanism implements pairwise interactions to dynamically aggregate long-range dependencies of feature representations according to the input content from various data modalities~\cite{vaswani2017attention}. TransUNet~\cite{chen2021transunet} adopts Transformer blocks to collect global correlations on top of ResNet backbone. UTNet~\cite{gao2021utnet} instead incorporates interleaved transformer blocks and convolution blocks for small medical dataset. MCTrans~\cite{ji2021multi} employs a Transformer Cross-Attention (TCA) module to collect context information from feature maps of different scales. However, these approaches are designed for single-view and can be sub-optimal for complex segmentation tasks due to the absence of considering semantic dependencies of different scales and views, which are critical for enhancing clinical lesion assessment.

In this paper, we propose TransFusion to merge divergent information from multiple views and scales, using powerful attention mechanisms for medical image segmentation. We employ convolutional layers for local feature extraction and develop strong attentive mechanisms to aggregate long-range correlation information from cross-scale and cross-view feature representations. To gather long-range dependencies among image views, we propose the Divergent Fusion Attention (DiFA) module to achieve cross-view context modeling and semantic dependency mining. Further, the Multi-Scale Attention (MSA) module is designed to collect global correspondence of multi-scale feature representations to ensure feature consistency at different levels of the pyramidal architecture. We evaluate TransFusion on the Multi-Disease, Multi-View \& Multi-Center
Right Ventricular Segmentation (M\&Ms-2) challenge cohort. TransFusion demonstrates leading performance against the state-of-the-art methods and holds the promise for a wide range of medical image segmentation tasks. To the best of our knowledge, our method is the first work to apply Transformer to multi-view medical image segmentation tasks, which also shows potential in wide applications for unaligned or multi-modality data.

\section{Methods}

\begin{figure} [t]
\begin{center}
\includegraphics[width=1\linewidth]{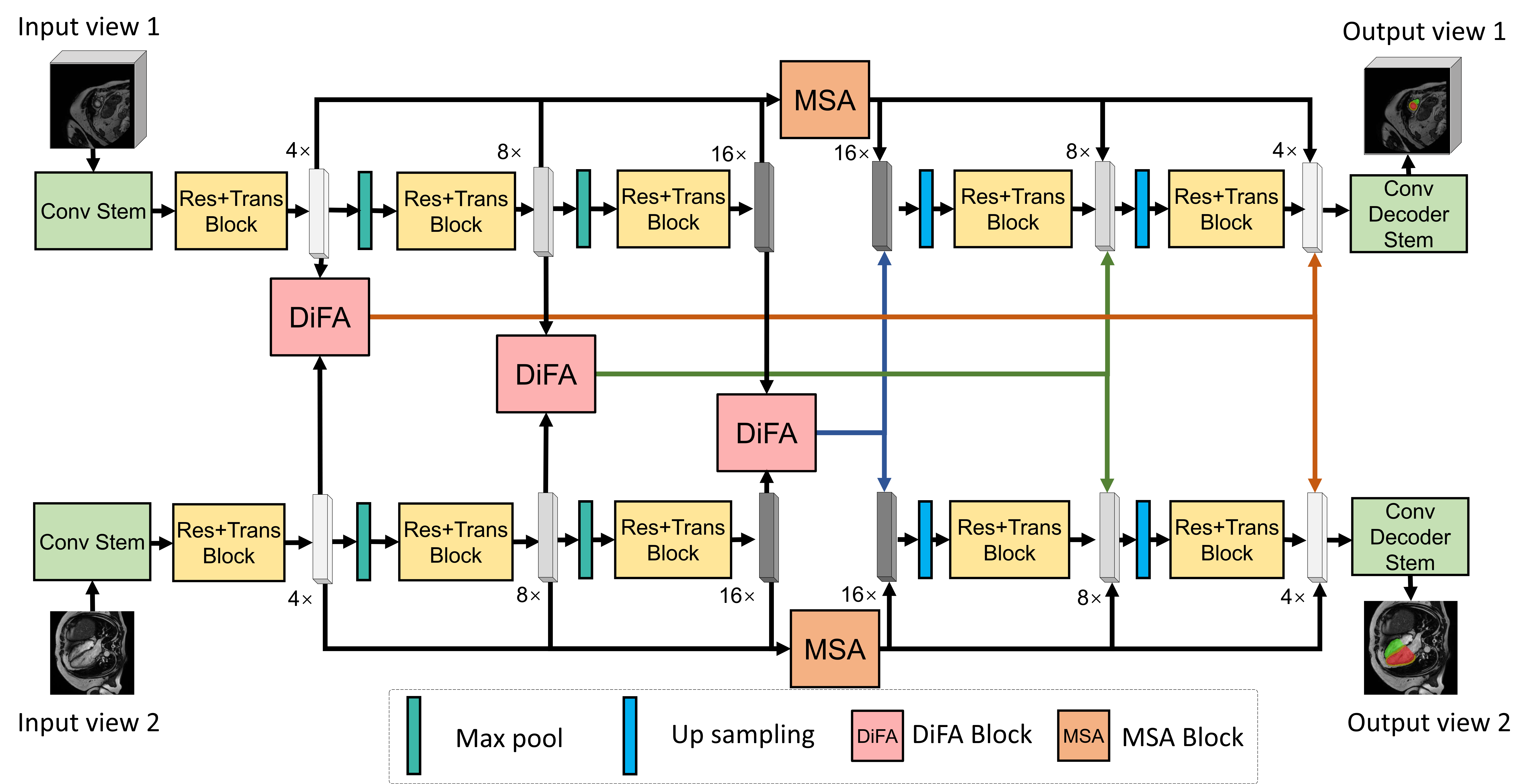}
\end{center}
   \caption{Overview of the Transfusion architecture. The Divergent Fusion Attention (DiFA) module enables rich cross-view context modeling and semantic dependency mining, capturing long-range dependencies between unaligned data from different image views. For each view, the proposed Multi-Scale Attention (MSA) module collects global correspondence of multi-scale feature representations.}
\label{framework}
\end{figure}

Fig. \ref{framework} highlights the proposed TransFusion framework. The inputs to the model are images from $M$ views. As the size, modality, and even dimension can be different among views, TransFusion has a sub-network for each view, where each sub-network can be an arbitrary network structure designed for any input image modality. Further, TransFusion has the ability to find correspondences and align features from heterogeneous inputs. In our experiments, 3D short-axis cMRI and 2D long-axis cMRI are used as multi-view inputs. For the purpose of efficiency, 3D UTNet \cite{gao2021utnet} and 2D UTNet are used as the backbone of sub-networks for the two inputs, which consist of interleaved Residual blocks and efficient Transformer blocks. As seen in Fig. \ref{framework}, for joint modeling of multi-view data and fusing multi-scale features, we apply the Divergent Fusion Attention (DiFA) block and multi-scale attention block (MSA) on the middle- and high-level token maps (Fig.~\ref{framework} grey modules). The three outputs of DiFA at different scales are further folded back to their corresponding scales of decoders as indicated by the blue, green and orange arrows. The convolution stem consists of multiple convolution layers and down-samplings to reduce the spatial resolution by $4\times$. The Res+Trans Block is the stack of a Residual Block and a Transformer Block (see Fig.~\ref{difa}). 


\subsection{Revisiting the Self-Attention Mechanism}

Transformers use the Multi-Head Self-Attention (MHSA) module for the modeling of data dependencies without considering their distance in different representation sub-spaces~\cite{bahdanau2014neural,kim2017structured,vaswani2017attention}. The outputs of multiple heads are concatenated through the Feed-forward Network (FFN). Given an input feature map $X \in {{\mathbb{R}}^{C \times H \times W}}$, where $C,H,W$ are the number of input channels, the spatial height and width, respectively (e.g. short and long axis cardiac MR images). The input $X$ is first projected to query $Q$, key $K$ and value $V$, through three linear projections. Here $Q,K,V \in {{\mathbb{R}}^{d \times H \times W}}$, where $d$ represents the embedding dimensions of each head. $Q,K,V$ are flattened and transposed into three token sequences with the same size $n\times d$ where $n=HW$. The output of self-attention layer can be denoted by a scaled dot-product as

\begin{equation}
{\rm{Attn(}}Q{\rm{,}}K{\rm{,}}V{\rm{) = softmax(}}\frac{{Q{K^T}}}{{\sqrt d }}{\rm{)V}},
\end{equation}
where ${\rm{softmax(}}\frac{{Q{K^T}}}{{\sqrt d }}{\rm{)}} \in {{\mathbb{R}}^{n \times n}}$ is the attention matrix, which computes the pair-wise similarity between each token of queries and keys, and then is used as the weights to collect context information from the values. This mechanism allows the modeling of long-range data dependencies for global feature aggregation.  
\subsection{Divergent Fusion Attention Module (DiFA)}

\begin{figure*} [t]
\begin{center}
\includegraphics[width=1\linewidth]{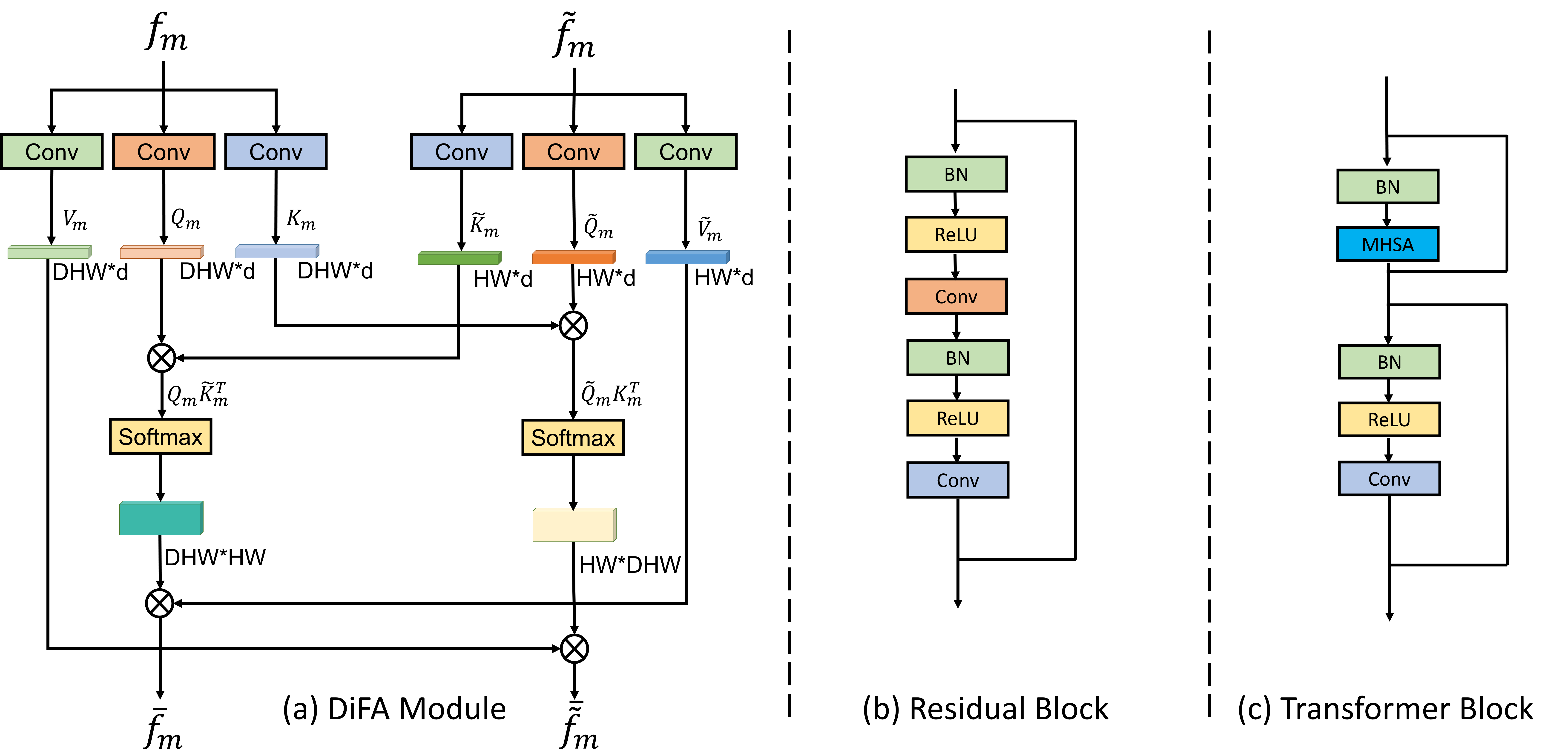}
\end{center}
   \caption{Illustration of the (a) Divergent Fusion Attention Module (DiFA), (b) Residual block with pre-activation and (c) Transformer block.}
\label{difa}
\end{figure*}

The self-attention mechanism offers global feature aggregation that is built upon single input features. For multi-view tasks, we seek to jointly model inputs from different views. We extend the self-attention to be applicable for multi-view fusion and thus introduce the Divergent Fusion Attention (DiFA) mechanism. The proposed DiFA aims to find correspondences in unaligned data and complement the missing information among multi-view images such as long and short axis cardiac images. For instance, the high-resolution information of the long-axis cardiac image can be used to complement the base and the apex the ventricles on the short-axis image. We take 3D short-axis and 2D long-axis cMRI inputs as an example shown in Fig. \ref{difa}.

Given token maps $f_0, f_1, ..., f_M \in \mathbb{R}^{C\times N_m}$ from $M$ views, where $N_m$ represents the token number in a single view (e.g. $N_m=H_m\times W_m$ for 2D input or $N_m=D_m\times H_m\times W_m$ for 3D input, $m\in \{0, ..., M\}$). All token maps $f_m$ are linearly projected into ${Q_m}$, ${K_m}$, ${V_m} \in {{\mathbb{R}}^{d \times N_m }}$. The main idea is to  enhance a specific view representation with context information from all non-target views. We first concatenate all non-target keys and values as $\tilde K_m,\tilde V_m \in {{\mathbb{R}}^{d \times \sum_{i=0,i\ne m}^{M} {{N_i}} }}$. The 
attention matrix is further computed by the target query $Q_m$ and the concatenated non-target embedding ${\tilde K_m}$. Then the updated $\bar{f}_m$ is: 

\begin{equation}
{{{ \bar{f}}_m} = \rm{DiFA(}}Q_m{\rm{,}}\tilde K_m,\tilde V_m{\rm{) = softmax(}}\frac{{Q_m{{\tilde {K}_m}^T}}}{{\sqrt d }}{\rm{)}}\tilde V_m
\end{equation}

We then repeat DiFA for all other views. In this way, the divergent context information from different views are fused in high-dimensional spaces to refine the initial representation from the target view only. Notably, as the inputs from different views are not aligned, the absolute or relative positional encoding are not used in our DiFA module. Further, the design of DiFA is not limited to multi-view images, but applied all data that is not well aligned, such as different modalities from 2D and 3D data.

\subsection{Multi-Scale Attention (MSA) Module}

To fuse the multi-scale representations in the hierarchical backbone, we introduce the Multi-Scale Attention (MSA) module to learn this contextual dependencies of inter-scale features. In practice, given the $m$-th view input, the encoder hierarchically extract features and obtain multiple-level feature $f_m^l$, where $l$ denotes the level of down-sampling. Then the output features $f_m^l \in \mathbb{R}^{C^l \times N_m^l}$ for every level are flattened and concatenated as the input of a Transformer block, which consists of a Multi-Head Self-Attention module and a Feed-forward Network. Following the feature extraction using CNN and Self-Attention, the MSA block is applied to fuse scale divergence. A single MSA layer is formulated as:

\begin{equation}
\bar f_m^0, \cdots ,\bar f_m^L = {\rm{FFN}}({\rm{Attn}}({\rm{Cat}}(f_m^0, \cdots ,f_m^L)))
\end{equation}
where ${{f}}_m^l = {\rm{Attn(}}Q_m^l{\rm{,}}K_m^l{\rm{,}}V_m^l{\rm{)}}$ denotes the flattened output of the Self-Attention Encoder layer at level $l$. The feature map of MSA is then folded back to the decoder at corresponding scales for further interaction and prediction.

\section{Experiments}
\subsection{Datasets and Settings}
The proposed TransFusion is systematically evaluated on the Multi-Disease, Multi-View \& Multi-Center
Right Ventricular Segmentation in Cardiac MRI (M\&Ms-2) challenge cohort~\cite{campello2021multi}. The M\&Ms-2 challenge cohort contains 160 scans, which were collected from subjects with different RV and LV pathologies in three clinical centers from Spain using three different magnetic resonance scanner vendors (Siemens, General Electric, and Philips). The acquired cine MR images were delineated by experienced clinical doctors, including the left ventricle (LV), right ventricle (RV) and the left ventricular myocardium (Myo). The whole cine MR sequences with varied number of frames are provided, but only ED and ES frames are labeled in both short-axis and long-axis views. 
Note that we treat the stack of short-axis images as 3D data, while the long-axis four-chamber view as 2D data. The training set includes subjects of normal, dilated left ventricle (DLV), hypertrophic cardiomyopathy (HC), congenital arrhythmogenesis (CA), tetralogy of fallot (TF) and interatrial comunication (IC).

We randomly shuffle the 160 samples and evaluate all models with 5-fold cross validation. All models are trained with Dice loss and focal loss, with batch size 32 and Adam optimizer for 300 epochs. The learning rate is 0.001 and decayed linearly by 99\% per epoch. Models are trained with eight Quadro RTX 8000 GPUs in PyTorch framework. For data preprocessing, all images are resampled to 1.25 mm in spacing. Data augmentation strategies are utilized, including random histogram matching, rotation, shifting, scaling, elastic deformation and mirroring. Dice score and Hausdorff distance are used as evaluation metrics.

\begin{table*}[tb]
\centering
\caption{ Comparison of methods in terms of Dice and Hausdorff Distance. (T) indicates the target view of segmentation. The bold texts mark the best performance. The methods marked with stars indicate they use multi-view inputs.}
\label{1}

\begin{tabular}{l|cc|cccc|cccc}
\hline
\multirow{2}{*}{Methods} & \multicolumn{2}{c|}{Input type}& \multicolumn{4}{c|}{Dice($\%$)} & \multicolumn{4}{c}{HD($mm$)} \\ \cline{2-11} 
              & SA(T) & LA & LV     & RV     & Myo &Avg    & LV     & RV     & Myo       &Avg  \\ \hline
UNet        & \checkmark &  - &  87.02 & 88.85 & 79.07 & 84.98  &  13.78 & 12.10    & 12.23 &12.70 \\
ResUNet~                                & \checkmark &  - &  87.98  & 89.63 & 79.28  &85.63 &  13.80 &11.61 & 12.09  &12.50  \\
DLA               & \checkmark &  - & 88.27   &  89.88 & 80.23 &86.13 &  13.25  &10.84     & 12.31 &12.13   \\
InfoTrans*        & \checkmark &  \checkmark &  88.24 & 90.41 &  80.25 & 86.30 & 12.41 & 10.98    &12.83  &12.07   \\
rDLA*         & \checkmark &  \checkmark &  88.64 & 90.28 &  80.78 & 86.57 & 12.74 & 10.31    &12.49   &11.85  \\
\hline
TransUNet   & \checkmark &  -  & 87.91   & 88.69  &  78.67 &85.09 & 13.80   & 10.29   & 13.45 &12.51   \\
MCTrans              & \checkmark &  - &  88.52 & 89.90  & 80.08 &86.17 & 12.29  &  9.92 & 13.28 &11.83\\
MCTrans*~            & \checkmark &  \checkmark & 87.79   & 89.22  & 79.37 & 85.46 &  \textbf{11.28} & 9.36 & 13.84 &11.49\\
UTNet~       & \checkmark &  - & 87.52   & 90.57  & 80.20 &86.10 &  12.03 & 9.78 &13.72 &11.84 \\
UTNet*~              & \checkmark &  \checkmark & 87.74   & 90.82 &  80.71 &86.42 & 11.79  & 9.11  & 13.41 &11.44 \\
Proposed*        & \checkmark &  \checkmark & \textbf{89.52}   & \textbf{91.75}  & \textbf{81.46} &\textbf{87.58}  & 11.31  & \textbf{9.18} & \textbf{11.96}  &\textbf{10.82} \\ \hline \hline

\multirow{2}{*}{Methods} & \multicolumn{2}{c|}{Input type}& \multicolumn{4}{c|}{Dice($\%$)} & \multicolumn{4}{c}{HD($mm$)} \\ \cline{2-11} 
              & SA & LA(T) & LV     & RV     & Myo  &Avg   & LV     & RV     & Myo   &Avg      \\ \hline
UNet       &  - & \checkmark  & 87.26 & 88.20 & 79.96 & 85.14 & 13.04 & 8.76 & 12.24 &11.35\\
ResUNet~      &  - & \checkmark &    87.61 & 88.41 & 80.12 &85.38& 12.72 &8.39 & 11.28&10.80 \\
DLA                 &  - & \checkmark &  88.37 & 89.38 & 80.35&86.03 & 11.74 & 7.04 & 10.79 &9.86\\
InfoTrans*        & \checkmark &  \checkmark &  88.21 & 89.11 &  80.55 &85.96 &  12.47 & 7.23    &10.21   &9.97  \\
rDLA*         & \checkmark &  \checkmark &   88.71   & 89.71 & 81.05 &86.49& 11.12 & 6.83 & 10.42&9.46\\
\hline
TransUNet   &  - & \checkmark     & 87.91 & 88.23 & 79.05 &85.06 & 12.02 & 8.14 & 11.21  &10.46  \\
MCTrans       &  - & \checkmark &  88.42 & 88.19 & 79.47&85.36 & 11.78 & 7.65 & 10.76 &10.06 \\
MCTrans*~     & \checkmark &  \checkmark & 88.81 & 88.61 & 79.94 &85.79 & 11.52 & 7.02 & 10.07 &9.54\\
UTNet       &  - & \checkmark & 86.93 & 89.07 & 80.48 &85.49 &   11.47 & 6.35  & 10.02 &9.28\\ 
UTNet*~               & \checkmark &  \checkmark &   87.36  & 90.42 & 81.02 &86.27& 11.13 & 5.91 & 9.81&8.95\\
Proposed*                               & \checkmark &  \checkmark & \textbf{89.78}   & \textbf{91.52}  & \textbf{81.79} &\textbf{87.70} & \textbf{10.25}  & \textbf{5.12} & \textbf{8.69} & \textbf{8.02}  \\ \hline 

\end{tabular}
\end{table*}

\begin{figure}[tb]
\setlength{\belowcaptionskip}{-7pt}
     \centering
     \begin{subfigure}[b]{0.44\textwidth}
         \centering
         \includegraphics[width=1\textwidth]{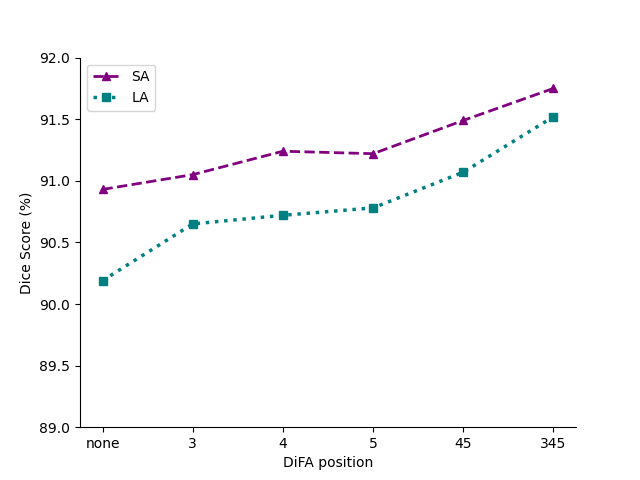}
         \caption{}
     \end{subfigure}
     \begin{subfigure}[b]{0.44\textwidth}
         \centering
         \includegraphics[width=1\textwidth]{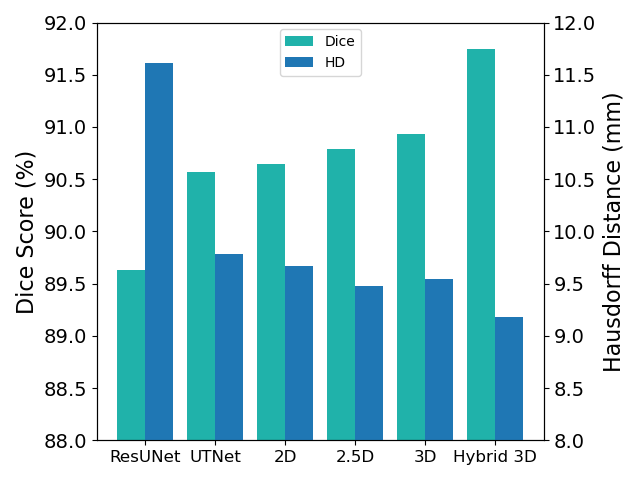}
         \caption{}
         
     \end{subfigure}
\caption{Ablation studies. (a) Effect of different DiFA settings. The number indicates which layer DiFA is performed. (b) Effect of network mode and comparison with two major baselines ResUNet and UTNet.}
\label{as}
\end{figure}

\subsection{Results}
Table \ref{1} compares the performance of TransFusion with several state-of-the-art approaches. ResUNet applies residual blocks as UNet building blocks~\cite{ronneberger2015u}. MCTrans~\cite{ji2021multi} introduces a cross-attention block between the encoder and decoder to gather cross-scale dependencies of the feature maps. Refined DLA (rDLA)~\cite{liu2021refined} bases its backbone on a leading CNN architecture, Deep Layer Aggregation (DLA)~\cite{yu2018deep}, and aggregates context information from cross-view through a refinement stage. InfoTrans*~\cite{li2021right} similarly aggregates information from cross-views using information transition. The result shows our TransFusion outperforms the other methods in terms of both short- and long-axis cardiac MR segmentation.

\begin{figure*} [t]
\begin{center}
\includegraphics[width=0.8\linewidth]{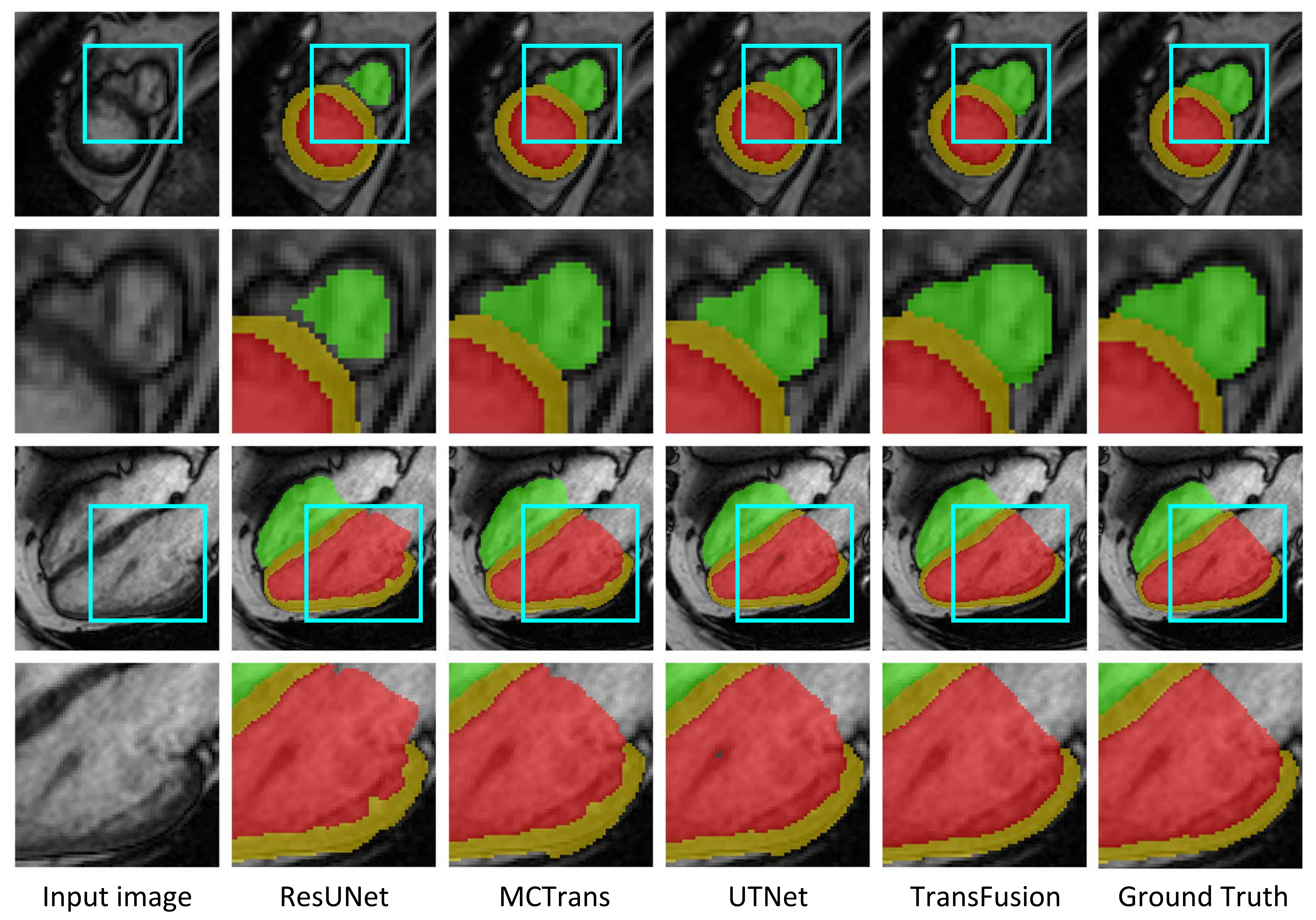}
\end{center}
  \caption{Segmentation results of short-(top) and long-axis (bottom) images.}
\label{vis}
\end{figure*}

We provide a further comparison by applying the proposed DiFA module to these existing baselines for multi-view segmentation tasks as indicated as starred methods in Table \ref{1}. The starred MCTrans and UTNet perform better than the single-view networks due to the mutual feature aggregation of the proposed DiFA. Fig. \ref{vis} further shows that Transfusion displays leading performance as compared to other competitive approaches in multi-view segmentation results.

\subsection{Ablation Studies}

\subsubsection{Sensitivity to the DiFA setting.}
We evaluate the performance of the Divergent Fusion Attention (DiFA) modules of TransFusion by the segmentation accuracy. As seen in Fig. \ref{as}(a), the $x$-axis number indicates the level where DiFAs are located, e.g., '45' means two DiFAs are performed respectively in the fourth and fifth level of the encoder. As the level goes up, the DiFA module is able to collect more detailed and divergent information from the other views. Note that the curve saturates when adding to the fourth level. However, when adding DiFA to multiple levels, we recognize that the learned cross-view prior helps construct identified context dependencies and improve the performance constantly with limited additional computational cost.

\subsubsection{Analysis of the network mode.}
Fig. \ref{as}(b) shows the performance using different network modes. The 2.5D mode combines two neighboring slices of short-axis sample and forms a three-channel input, allowing a local spatial information fusion for segmentation. The proposed TransFusion applies a hybrid 3D + 2D mode and outperforms other baselines, which is attributed to the alignment of spatial information between short- and long-axis data. For each short-axis sample, as more through-plane textures are included to construct the cross-view dependencies with the corresponding long-axis image, our TransFusion can better build up global context dependencies through the DiFA modules.

\section{Conclusion}
We have proposed TransFusion, a powerful Transformer architecture, to merge critical cross-view information towards enhanced segmentation performance using convolutional layers and powerful attentive mechanisms for medical images. The proposed Multi-Scale Attention (MSA) and Divergent Fusion Attention (DiFA) modules allow rich cross-scale and cross-view context modeling and semantic dependency mining, effectively addressing the issues of capturing long-range dependencies within as well as between different scales and views. This strong ability further opens up new perspectives for applying TransFusion on more downstream view-fusion tasks in medical imaging and computer vision. 

\bibliographystyle{splncs04}
\bibliography{paper1860}

\begin{thebibliography}{10}
\providecommand{\url}[1]{\texttt{#1}}
\providecommand{\urlprefix}{URL }
\providecommand{\doi}[1]{https://doi.org/#1}

\bibitem{bahdanau2014neural}
Bahdanau, D., Cho, K., Bengio, Y.: Neural machine translation by jointly
  learning to align and translate. arXiv preprint arXiv:1409.0473  (2014)

\bibitem{bernard2018deep}
Bernard, O., Lalande, A., Zotti, C., Cervenansky, F., Yang, X., Heng, P.A.,
  Cetin, I., Lekadir, K., Camara, O., Ballester, M.A.G., et~al.: Deep learning
  techniques for automatic mri cardiac multi-structures segmentation and
  diagnosis: is the problem solved? IEEE transactions on medical imaging
  \textbf{37}(11),  2514--2525 (2018)

\bibitem{campello2021multi}
Campello, V.M., Gkontra, P., Izquierdo, C., Mart{\'\i}n-Isla, C., Sojoudi, A.,
  Full, P.M., Maier-Hein, K., Zhang, Y., He, Z., Ma, J., et~al.: Multi-centre,
  multi-vendor and multi-disease cardiac segmentation: the m\&ms challenge.
  IEEE Transactions on Medical Imaging  \textbf{40}(12),  3543--3554 (2021)

\bibitem{cao2021swin}
Cao, H., Wang, Y., Chen, J., Jiang, D., Zhang, X., Tian, Q., Wang, M.:
  Swin-unet: Unet-like pure transformer for medical image segmentation. arXiv
  preprint arXiv:2105.05537  (2021)

\bibitem{chang2020soft}
Chang, Q., Yan, Z., Lou, Y., Axel, L., Metaxas, D.N.: Soft-label guided
  semi-supervised learning for bi-ventricle segmentation in cardiac cine mri.
  In: 2020 IEEE 17th International Symposium on Biomedical Imaging (ISBI). pp.
  1752--1755. IEEE (2020)

\bibitem{chang2022deeprecon}
Chang, Q., Yan, Z., Zhou, M., Liu, D., Sawalha, K., Ye, M., Zhangli, Q.,
  Kanski, M., Aref, S.A., Axel, L., Metaxas, D.: Deeprecon: Joint 2d cardiac
  segmentation and 3d volume reconstruction via a structure-specific generative
  method. arXiv preprint arXiv:2206.07163  (2022)

\bibitem{chen2021transunet}
Chen, J., Lu, Y., Yu, Q., Luo, X., Adeli, E., Wang, Y., Lu, L., Yuille, A.L.,
  Zhou, Y.: Transunet: Transformers make strong encoders for medical image
  segmentation. arXiv preprint arXiv:2102.04306  (2021)

\bibitem{dosovitskiy2020image}
Dosovitskiy, A., Beyer, L., Kolesnikov, A., Weissenborn, D., Zhai, X.,
  Unterthiner, T., Dehghani, M., Minderer, M., Heigold, G., Gelly, S., et~al.:
  An image is worth 16x16 words: Transformers for image recognition at scale.
  arXiv preprint arXiv:2010.11929  (2020)

\bibitem{gao2022multi}
Gao, Y., Zhou, M., Liu, D., Metaxas, D.: A multi-scale transformer for medical
  image segmentation: Architectures, model efficiency, and benchmarks. arXiv
  preprint arXiv:2203.00131  (2022)

\bibitem{gao2021utnet}
Gao, Y., Zhou, M., Metaxas, D.N.: Utnet: a hybrid transformer architecture for
  medical image segmentation. In: International Conference on Medical Image
  Computing and Computer-Assisted Intervention. pp. 61--71. Springer (2021)

\bibitem{ge2020automated}
Ge, C., Liu, D., Liu, J., Liu, B., Xin, Y.: Automated recognition of arrhythmia
  using deep neural networks for 12-lead electrocardiograms with fractional
  time--frequency domain extension. Journal of Medical Imaging and Health
  Informatics  \textbf{10}(11),  2764--2767 (2020)

\bibitem{hatamizadeh2022unetr}
Hatamizadeh, A., Tang, Y., Nath, V., Yang, D., Myronenko, A., Landman, B.,
  Roth, H.R., Xu, D.: Unetr: Transformers for 3d medical image segmentation.
  In: Proceedings of the IEEE/CVF Winter Conference on Applications of Computer
  Vision. pp. 574--584 (2022)

\bibitem{he2019effective}
He, X., Tan, C., Qiao, Y., Tan, V., Metaxas, D., Li, K.: Effective 3d humerus
  and scapula extraction using low-contrast and high-shape-variability mr data.
  In: Medical Imaging 2019: Biomedical Applications in Molecular, Structural,
  and Functional Imaging. vol. 10953, p. 109530O. International Society for
  Optics and Photonics (2019)

\bibitem{hu2020harnessing}
Hu, J.B., Guan, A., Zhangli, Q., Sayadi, L.R., Hamdan, U.S., Vyas, R.M.:
  Harnessing machine-learning to personalize cleft lip markings. Plastic and
  Reconstructive Surgery--Global Open  \textbf{8}(9S),  150--151 (2020)

\bibitem{hu2003vivo}
Hu, Z., Metaxas, D., Axel, L.: In vivo strain and stress estimation of the
  heart left and right ventricles from mri images. Medical image analysis
  \textbf{7}(4),  435--444 (2003)

\bibitem{ji2021multi}
Ji, Y., Zhang, R., Wang, H., Li, Z., Wu, L., Zhang, S., Luo, P.: Multi-compound
  transformer for accurate biomedical image segmentation. In: International
  Conference on Medical Image Computing and Computer-Assisted Intervention. pp.
  326--336. Springer (2021)

\bibitem{kim2017structured}
Kim, Y., Denton, C., Hoang, L., Rush, A.M.: Structured attention networks.
  arXiv preprint arXiv:1702.00887  (2017)

\bibitem{li2021right}
Li, L., Ding, W., Huang, L., Zhuang, X.: Right ventricular segmentation from
  short-and long-axis mris via information transition. arXiv preprint
  arXiv:2109.02171  (2021)

\bibitem{liu2020dispersion}
Liu, D., Ge, C., Xin, Y., Li, Q., Tao, R.: Dispersion correction for optical
  coherence tomography by the stepped detection algorithm in the fractional
  fourier domain. Optics express  \textbf{28}(5),  5919--5935 (2020)

\bibitem{liu2021label}
Liu, D., Liu, J., Liu, Y., Tao, R., Prince, J.L., Carass, A.: Label super
  resolution for 3d magnetic resonance images using deformable u-net. In:
  Medical Imaging 2021: Image Processing. vol. 11596, p. 1159628. International
  Society for Optics and Photonics (2021)

\bibitem{liu2019dispersion}
Liu, D., Xin, Y., Li, Q., Tao, R.: Dispersion correction for optical coherence
  tomography by parameter estimation in fractional fourier domain. In: 2019
  IEEE International Conference on Mechatronics and Automation (ICMA). pp.
  674--678. IEEE (2019)

\bibitem{liu2021refined}
Liu, D., Yan, Z., Chang, Q., Axel, L., Metaxas, D.N.: Refined deep layer
  aggregation for multi-disease, multi-view \& multi-center cardiac mr
  segmentation. In: International Workshop on Statistical Atlases and
  Computational Models of the Heart. pp. 315--322. Springer (2021)

\bibitem{petitjean2011review}
Petitjean, C., Dacher, J.N.: A review of segmentation methods in short axis
  cardiac mr images. Medical image analysis  \textbf{15}(2),  169--184 (2011)

\bibitem{remedios2021joint}
Remedios, S.W., Han, S., Dewey, B.E., Pham, D.L., Prince, J.L., Carass, A.:
  Joint image and label self-super-resolution. In: International Workshop on
  Simulation and Synthesis in Medical Imaging. pp. 14--23. Springer (2021)

\bibitem{ronneberger2015u}
Ronneberger, O., Fischer, P., Brox, T.: U-net: Convolutional networks for
  biomedical image segmentation. In: International Conference on Medical image
  computing and computer-assisted intervention. pp. 234--241. Springer (2015)

\bibitem{tian2018cr}
Tian, Y., Peng, X., Zhao, L., Zhang, S., Metaxas, D.N.: Cr-gan: learning
  complete representations for multi-view generation. arXiv preprint
  arXiv:1806.11191  (2018)

\bibitem{vaswani2017attention}
Vaswani, A., Shazeer, N., Parmar, N., Uszkoreit, J., Jones, L., Gomez, A.N.,
  Kaiser, {\L}., Polosukhin, I.: Attention is all you need. In: Advances in
  neural information processing systems. pp. 5998--6008 (2017)

\bibitem{vigneault2018omega}
Vigneault, D.M., Xie, W., Ho, C.Y., Bluemke, D.A., Noble, J.A.: $\omega$-net
  (omega-net): fully automatic, multi-view cardiac mr detection, orientation,
  and segmentation with deep neural networks. Medical image analysis
  \textbf{48},  95--106 (2018)

\bibitem{wang2017multi}
Wang, S., Zhou, M., Gevaert, O., Tang, Z., Dong, D., Liu, Z., Jie, T.: A
  multi-view deep convolutional neural networks for lung nodule segmentation.
  In: 2017 39th Annual International Conference of the IEEE Engineering in
  Medicine and Biology Society (EMBC). pp. 1752--1755. IEEE (2017)

\bibitem{xia2020uncertainty}
Xia, Y., Yang, D., Yu, Z., Liu, F., Cai, J., Yu, L., Zhu, Z., Xu, D., Yuille,
  A., Roth, H.: Uncertainty-aware multi-view co-training for semi-supervised
  medical image segmentation and domain adaptation. Medical Image Analysis
  \textbf{65},  101766 (2020)

\bibitem{yu2018deep}
Yu, F., Wang, D., Shelhamer, E., Darrell, T.: Deep layer aggregation. In:
  Proceedings of the IEEE conference on computer vision and pattern
  recognition. pp. 2403--2412 (2018)

\bibitem{zhangli2022region}
Zhangli, Q., Yi, J., Liu, D., He, X., Xia, Z., Tang, H., Wang, H., Zhou, M.,
  Metaxas, D.: Region proposal rectification towards robust instance
  segmentation of biological images. arXiv preprint arXiv:2203.02846  (2022)

\bibitem{zhao2019applications}
Zhao, C., Shao, M., Carass, A., Li, H., Dewey, B.E., Ellingsen, L.M., Woo, J.,
  Guttman, M.A., Blitz, A.M., Stone, M., et~al.: Applications of a deep
  learning method for anti-aliasing and super-resolution in mri. Magnetic
  resonance imaging  \textbf{64},  132--141 (2019)

\end{thebibliography}
%





\end{document}